\documentclass[aps,prd,superscriptaddress,nofootinbib,11pt]{revtex4}
\usepackage[english]{babel}
\usepackage[utf8]{inputenc}
\usepackage{graphicx}   
\usepackage{slashed}
\usepackage{epstopdf}
\usepackage{verbatim}   
\usepackage{color}      
\usepackage{subfigure}  
\usepackage{multirow}
\usepackage{hyperref}   
\usepackage{float}
\usepackage{epsfig,rotating}
\usepackage{amsmath,amssymb}
\usepackage{dsfont}
\usepackage{slashed}
\restylefloat{table}
\numberwithin{equation}{section}

\newcommand{\vx}{\vec{x}}

\newcommand{\vk}{\vec{k}}

\newcommand{\op}{\mathcal{O}_\chi}

\newcommand{\be}{\begin{equation}}
\newcommand{\ee}{\end{equation}}
\newcommand{\bea}{\begin{eqnarray}}
\newcommand{\eea}{\end{eqnarray}}

\newcommand{\ket}[1]{|#1\rangle}
\newcommand{\bra}[1]{\langle#1|}

\begin{document}
\title{ Chern Simons condensate from misaligned axions.}

\author{Shuyang Cao}
\email{shuyang.cao@pitt.edu} \affiliation{Department of Physics, University of Pittsburgh, Pittsburgh, PA 15260}
\author{Daniel Boyanovsky}
\email{boyan@pitt.edu} \affiliation{Department of Physics, University of Pittsburgh, Pittsburgh, PA 15260}

 \date{\today}

\begin{abstract}

We   obtain the non-equilibrium condensate of the Chern Simons density induced by a misaligned homogeneous coherent axion field in linear response.  The Chern-Simons dynamical susceptibility is simply related to  the axion self-energy,  a result that is valid to leading order in the axion coupling but to all orders in the couplings of the gauge fields to other fields within or beyond the standard model except   the axion. The induced Chern-Simons density requires renormalization which is achieved by vacuum subtraction. For ultralight axions of mass $m_a$ coupled to electromagnetic fields with coupling $g$,    the renormalized high  temperature  contribution post-recombination  is $\langle \vec{E}\cdot\vec{B}\rangle(t) =  -\frac{g\,\pi^2\,T^4}{15}    \,\overline{a}(t)+ \frac{g \,m^2_a\,T}{16\,\pi}\,\dot{\overline{a}}(t) $ with $\overline{a}(t)$ the  dynamical homogeneous axion condensate.  We conjecture that emergent axion-like quasiparticle excitations in condensed matter systems may be harnessed to probe cosmological axions and the Chern-Simons condensate. Furthermore, it is  argued that a misaligned axion can also induce  a non-abelian Chern-Simons condensate of similar qualitative form, and can also ``seed'' chiral symmetry breaking and induce a neutral pion condensate after the QCD phase transition.

\end{abstract}


\maketitle

\section{Introduction}\label{sec:intro}

The strong (CP) problem in quantum chromodynamics (QCD) motivated the proposal of a new pseudoscalar particle beyond the standard model, the axion, as a possible solution\cite{PQ,weinaxion,wil} by elevating an (CP) violating angle to a dynamical field. Such field   may be produced non-thermally in the Early Universe, for example by a misalignment mechanism in which an initial axion coherent condensate is produced out of equilibrium and evolves towards the minimum of its (effective) potential. Such axion field has also been recognized as a potentially viable  cold dark matter candidate\cite{pres,abbott,dine}. Pseudoscalar particles with properties similar to the QCD axion can also be accommodated   within suitable extensions  beyond the standard model, collectively referred to as  axion-like-particles (ALP),  which can also be   dark matter candidates\cite{banks,ringwald,marsh,sikivie1,sikivie2}, in particular as compelling candidates for ultra light dark matter\cite{fuzzy,uldm}. Furthermore, a dynamical misaligned  axion coherent condensate could also be a dark energy candidate in the form of a quintessence field whose slow dynamical evolution towards an equilibrium minimum would induce   an accelerated cosmological expansion phase\cite{quint}.

Constraints on the mass and couplings of ultra light (ALP)\cite{marsh,sikivie1,sikivie2,banik} are being established  by various observations and experiments ranging from astrophysical phenomena to table-top experiments\cite{cast,admx,graham}. There are two important features that characterize (ALP), i) a misalignment mechanism results in coherent oscillations of the expectation value of the (ALP) field which gives rise to its contribution to the energy density as a cold dark matter component\cite{pres,abbott,dine,marsh,sikivie1,sikivie2,turner}, ii) its pseudoscalar nature leads to an interaction between the (ALP) and photons  or  gluons via pseudoscalar composite operators of gauge fields, such as $F_{\mu\nu}\widetilde{F}^{\mu \nu}$ in the case of the (ALP)-photon interaction and $G^{\mu\nu;b}\widetilde{G}_{\mu\nu;b}$ in the case of gluons. We refer to these operators as Chern-Simons terms which are total surface terms. Such couplings were originally studied in ref.\cite{carroll} within the context of parity and Lorentz violating extensions of the standard model and early limits on these couplings were established from birefringence effects, namely different dispersion relations for different polarizations, and the rotation of the plane of polarization  from astrophysical sources. A telltale feature of birefringence from the electromagnetic coupling to axions is that the polarization rotation angle is  frequency independent\cite{harari,finelli}, which differentiates it from the more familiar Faraday effect resulting from the presence of magnetic fields in the  astrophysical plasma. Optical properties of axion backgrounds have been discussed in ref.\cite{mcdonald}, further electromagnetic signatures of axion electrodynamics were studied in refs.\cite{beutter,arza}, and photon production from parametric amplification of a misaligned axion condensate was studied in ref.\cite{dashin}.
 Analysis  of evidence for parity violating effects in the  Planck 2018 polarization data of the cosmic microwave background (CMB) anisotropies, revealed a non-vanishing cosmic birefringence angle at the  $2.4\sigma$ level\cite{minami} and more recently a combined analysis of WMAP and Planck polarization data revealed hints of isotropic cosmic birefringence at the $3\sigma$ level\cite{murai,komatsu,pala,eskilt}. These tantalizing hints may be a signal of cosmological axions.

  Axions may also play a role in condensed matter physics, possibly  as emergent quasiparticles in  topological insulators where magnetic fluctuations couple to electromagnetism just like axions\cite{wilczekaxion,wang,narang}, as axionic charge density waves in Weyl semimetals\cite{gooth,yu}, or as an emergent axion response in multilayered metamaterials with tunable couplings\cite{wilczek} or in multiferroics\cite{bala}. The measurement of an emergent dynamic axion field in chromia has been reported in ref.\cite{binek},  therefore, condensed matter systems may very well provide an experimental platform to test the main aspects of axion electrodynamics which may complement and bolster the case for axions in cosmology. Hence,  the study of axion (electro) dynamics is of timely interdisciplinary relevance. In this article we suggest that axion-like quasiparticles in condensed matter systems mix  with the cosmological axion, therefore topological insulators, Weyl semimetals or metamaterials  may provide   experimental  platforms to probe the cosmological axion and    the Chern-Simons condensate.

\subsection{Motivation and objectives:}\label{subsec:objectives}

The possibility of an axion or (ALP) being the dark matter and or dark energy candidate with a hallmark signature of frequency independent cosmic birefringence motivates a study of its non-equilibrium evolution when coupled to standard model degrees of freedom. Recently, refs.\cite{shuyang} implemented methods borrowed from non-equilibrium quantum field theory, namely the in-in Keldysh-Schwinger formulation,  and the theory of quantum open systems to study the non-equilibrium dynamics of axion-like  particles coupled to a bath in thermal equilibrium. These studies focused on the damping of a coherent misaligned condensate   as a consequence of its decay into photons from its coupling to electromagnetic fields via the Chern-Simons density,   the concomitant thermalization   of the axion fluctuations with the (CMB) photons yielding a mixed dark matter scenario, and an assessment of the  time scales of decoherence and entropy production. The Schwinger-Keldysh formulation of non-equilibrium quantum field theory is   suited to obtain the causal equations of motion of the axion field in presence of a heat bath. These were obtained in ref.\cite{shuyang}   and shown to be stochastic, of the Langevin type with a Gaussian noise and  include the retarded self-energy. The self-energy describes the damping of axion oscillations as a consequence of decay into the bath degrees of freedom (radiation reaction), and damping and noise are related by the quantum fluctuation dissipation relation, a consequence of which is the thermalization of axion fluctuations with the heat bath.  An alternative method based on the quantum master equation confirms these results\cite{shuyang} and  unequivocally shows that damping, thermalization and decoherence are directly related and occur on similar time scales.

Motivated by the confluence of interest on non-equilibrium axion dynamics both in cosmology and in condensed matter physics, in this article we study   the emergence of a Chern-Simons (topological) condensate as a consequence of the non-equilibrium dynamics of a misaligned axion macroscopic coherent condensate. A  Chern-Simons condensate would be manifest as a  non-vanishing expectation value of the abelian Chern-Simons density $\vec{E}\cdot \vec{B}$ in the non-equilibrium density matrix that describes the heat bath and the dynamical misaligned axion condensate. This is distinctly  different from the classical treatments of axion electrodynamics studied in refs.\cite{mcdonald,harari,beutter,arza}, and to the best of our knowledge such study has not been previously undertaken. Furthermore, the possibility of testing axion electrodynamics in condensed matter systems, such as topological insulators and Weyl semimetals, bolsters the case for studying  the emergence of Chern-Simons condensates as an intrinsic, fundamental  non-equilibrium aspect of axion physics with interdisciplinary relevance.

To this aim we implement the theory of linear response, ubiquitous in many body physics\cite{fetter}, to obtain the Chern-Simons condensate \emph{induced} by a misaligned axion condensate. While this study is focused on studying the emergence of a Chern-Simons condensate in Minkowski space time, as an initial step towards a more complete understanding within the context of an expanding cosmology, the basic concepts are expected to translate to cosmology qualitatively, but with quantitative differences in the time evolution. Such study awaits the consistent extrapolation of the linear response treatment to the realm of an expanding cosmology.

This article is structured as follows: in section (\ref{sec:cs-condensate}) we define   various models wherein an axion field is coupled to generic composite pseudoscalar operators $\mathcal{O}$ describing degrees of freedom within (or beyond)  the standard model, and obtain an exact relation between the induced condensate $\langle \mathcal{O} \rangle$ and the dynamical expectation value of the axion field.
Linear response theory is implemented to  obtain   the induced non-equilibrium expectation value $\langle \mathcal{O} \rangle$   to leading order in the axion coupling, and  introduce the concept of the dynamical susceptibility, namely  the response kernel   that relates $\langle \mathcal{O} \rangle$ to the coherent misaligned axion condensate.  In this section we establish one of the main results: the dynamical susceptibility is directly and simply related to the axion self-energy. In section (\ref{sec:CS}), we apply these results to obtain the non-equilibrium expectation value $\langle \vec{E}\cdot \vec{B} \rangle$, namely the Chern-Simons condensate,  in axion electrodynamics. In this section we show that this condensate features ultraviolet divergences proportional to the dynamical axion condensate, which acts as an explicit parity-symmetry breaking term, we also obtain the high temperature contributions to the Chern-Simons condensate for ultralight axions. In section (\ref{sec:emeraxion})   we argue that axion-like quasiparticles in condensed matter systems such as topological insulators or Weyl semimetals  mix with the cosmological axion via correlation functions of the Chern-Simons density, and that a Chern-Simons condensate acts as  a non-equilibrium driving term coupled to the emergent axion field. Therefore these experimentally available systems may be harnessed to probe the cosmological axion and the Chern-Simons condensate.
In section (\ref{sec:discussions}) we discuss important caveats in the cosmological setting, subtle renormalization aspects of the Chern-Simons condensate, and we argue on corresponding results for the non-abelian case, albeit with caveats. We also conjecture that a misaligned axion condensate \emph{may} induce a neutral pion condensate and seed chiral symmetry breaking during the QCD phase transition.
Section (\ref{sec:conclusions}) summarizes our conclusions.

\section{Condensate Induced by Axion-Like Fields}\label{sec:cs-condensate}
In this section, we discuss a general composite pseudoscalar operator $\mathcal{O}$ coupled to axion-like fields and show that such coupling implies that  a coherent condensate of the axion field  induces  a macroscopic condensate of the pseudoscalar operator, namely $\langle\mathcal{O}\rangle \neq 0$. We obtain a formal  exact relation between the expectation value of the axion field and that of the composite operator $\mathcal{O}$.  We then implement linear response to obtain an  explicit relation between these condensates to leading order in the axion coupling.

\subsection{Exact relation between the (ALP) condensate and $\langle \mathcal{O} \rangle$.}\label{subsec:exactrela}
We consider  an axion-like field $ a(x)$ coupled to  a composite pseudoscalar operator $\op(x)$ of generic fields $\chi(x)$. The Lagrangian density is
\be     \mathcal{L}[a,\chi] = \frac{1}{2}\,\partial_\mu a(x) \partial^\mu a(x) - \frac{1}{2}\,m^2_{0a}\,a^2(x) - g a(x)\,\op(x) + \mathcal{L}_{\chi}
    \label{lag} \ee where $m_{0a}$ is the bare axion mass, $\mathcal{L}_{\chi}$ is the Lagrangian density describing the fields $\chi$. Some examples of fields $\chi$ are electromagnetic fields with $\op(x)=\vec{E}(x)\cdot\vec{B}(x)$, gluon fields with $\op(x)=G^{\mu \nu,b}(x)\widetilde{G}_{\mu \nu,b}(x)$, where the tilde stands for  the dual of the gauge fields,  or fermionic fields with $\op(x)=i\overline{\Psi}(x)\gamma^5 \Psi(x)$. These fields could also be coupled to other degrees of freedom within or beyond the Standard Model, which are also all included in $\mathcal{L}_\chi$.

For gauge fields, the operators $\op(x)$ are the Chern-Simons terms and are a total surface term\cite{carroll}. In $3+1$ dimensions, these operators feature dimension $(mass)^4$ and the axion-like field $a(x)$ features dimension $(mass)$, which means they couple via non-renormalizable interactions whose coupling strength features dimension $(mass)^{-1}$. Therefore, the axion interacting locally with gauge fields via Chern-Simons terms must be interpreted as an effective field theory, whose validity is restricted to scales below a cutoff $\Lambda$. Furthermore, at finite temperature $T$, the validity of the effective field theory requires that $\Lambda \gg T$ so that high energy degrees of freedom are not thermally excited.   This observation will become relevant in the discussion of the induced Chern-Simons condensate in the next section.

The Heisenberg equation of motion for the axion field obtained from the Lagrangian density (\ref{lag}) is
\be
    \frac{\partial^2}{\partial t^2} \,  {a}(\vx,t)-\nabla^2  {a}(\vx,t)+m^2_{0a}\, {a}(\vx,t)= - g \, \mathcal{O}_{\chi}(\vx,t)  \,,\label{eomaxion1}
\ee
where the time evolution of any operator $\mathcal{A}$ in the full Heisenberg picture is
\begin{equation}
    \mathcal{A}(\vx,t)= e^{i H (t-t_0)}\,\mathcal{A}(\vx,t_0) \, e^{-i H (t-t_0)}
\end{equation}
with $H= H_{\chi}+H_a+H_i$ being the total Hamiltonian for the $\chi$ and axion fields and their interaction. The expectation value of any Heisenberg picture field operator $\mathcal{A}$ is obtained as  $\langle \mathcal{A}(\vx,t) \rangle = Tr(\mathcal{A}(\vx,t) \rho(t_0))$ where $\rho(t_0)$ is the normalized,  initial density matrix. In the Heisenberg picture the density matrix does not depend on time, therefore, taking expectation value of the Heisenberg field equation (\ref{eomaxion1}) yields
\begin{equation}
    \langle \op(\vx,t)\rangle =  -\frac{1}{g}\Bigg[\frac{\partial^2}{\partial t^2} \, \overline{a}(\vx,t)-\nabla^2 \overline{a}(\vx,t)+m^2_{0a}\,\overline{a}(\vx,t) \Bigg]
    \label{eqn:exactO}
\end{equation}
where $\overline{a}(\vx,t)$ is the expectation value of axion-like fields solutions of the Heisenberg equation of motion (\ref{eomaxion1}). This is an \emph{exact} relation   valid for arbitrary initial conditions and to all orders in the various couplings, however, eqn. (\ref{eqn:exactO}) by itself does not yield a closed expression for $\langle \mathcal{O}_{\chi}(\vx,t)\rangle$. This is because $\overline{a}(\vx,t)$ is the solution of the full equation of motion (\ref{eomaxion1}), which is not known a priori and  is obtained perturbatively in general. The exact relation (\ref{eqn:exactO}) becomes useful when the solution $\overline{a}(\vx,t)$ is obtained.

\subsection{Linear Response:}\label{subsec:LR}
We now formulate the general theory of linear response implemented to obtain a non-equilibrium expectation value of composite pseudoscalar operators coupled to axion-like fields as in the Lagrangian (\ref{lag}), relegating its specific application to the electromagnetic Chern-Simons term to next section.

   The Lagrangian density (\ref{lag}) describes several relevant couplings of axion-like fields to generic fields $\chi$. For $g=0$ these degrees of freedom   are assumed to be  described by a parity even thermal equilibrium density matrix $\rho_\chi$, consequently
  \be \mathrm{Tr}\, \op(x)\,\rho_\chi = 0\,. \label{zeroavO}\ee

  A misaligned axion-like   condensate is described by a \emph{classical} field
  $\overline{a}(\vx,t)$ corresponding to the expectation value of the axion-like field in a coherent state density matrix that describes the axion field\cite{shuyang}. Therefore we can decompose $a(x) = \overline{a}(\vx,t)+\widetilde{a}(x)$ where $\widetilde{a}(x)$ corresponds to the fluctuations of the axion field around the condensate and features vanishing expectation value in the axion density matrix. As envisaged in cosmology, the misaligned condensate $\overline{a}(\vx,t)$ is a macroscopic field, if it is a quintessence field driving cosmological expansion, it is homogeneous  within at least the Hubble scale. Therefore, neglecting the fluctuations $\widetilde{a}(x)$ (this is a mean field approximation) the interaction term in (\ref{lag}) is
  $\mathcal{L}_I = - g \overline{a}(\vx,t)\,\op(x)$, hence the pseudoscalar coupling to the axion field results in that the fields $\chi$ are coupled to an
  a \emph{c-number} external source $\overline{a}(\vx,t)$ with an explicit time dependence determined by the time evolution of the misaligned condensate. In presence of this classical source, the total  time dependent Hamiltonian for the $\chi$ fields is
  \be \widetilde{H}_{\chi}(t) = H_{\chi}+H_I(t)\,, \label{htilchi}\ee where $H_{\chi}$ is the Schroedinger picture Hamiltonian of the $\chi$ fields including coupling to other fields within or beyond the Standard Model except the axion field, and
  \be H_I(t) = g \int d^3 x \,\overline{a}(\vx,t)\,\op(\vx)\,,  \label{hint} \ee is the interaction Hamiltonian in the Schroedinger picture of the $\chi$-fields, but with $\overline{a}(\vx,t)$ playing the role of an ``external'' time dependent source term.
  In the Schroedinger picture the $\chi$-field density matrix evolves in time as
  \be \rho_{\chi}(t) = U(t,t_0)\,\rho_{\chi}(t_0)\,U^{-1}(t,t_0)\,,\label{rhochioft}\ee
  where the unitary time evolution operator $ U(t,t_0)$ obeys
  \be i\frac{d}{dt} U(t,t_0) = \widetilde{H}_{\chi}(t)\,  U(t,t_0)~~;~~  U(t_0,t_0)=1\,.  \label{timevolU}\ee The initial density matrix $\rho_{\chi}(t_0)$  is \emph{assumed} to describe  an   ensemble of the $\chi$ degrees of freedom in thermal equilibrium at temperature $T= 1/\beta$, namely
  \be \rho_{\chi}(t_0) = \frac{e^{-\beta H_{\chi}}}{\mathrm{Tr}e^{-\beta H_{\chi}}}\,,\label{rho0}\ee therefore
  \be e^{iH_{\chi} t_0}\, \rho_{\chi}(t_0) \, e^{-iH_{\chi} t_0}= \rho_{\chi}(t_0)\,,\label{equichi} \ee since in general $[\op(\vx),\rho_{\chi}(t_0)] \neq 0$ it follows that $\rho_{\chi}$ evolves in time out of equilibrium.

  Writing
  \be U(t,t_0)  = e^{-iH_{\chi} t}\, \mathcal{U}(t,t_0)\,e^{iH_{\chi} t_0}  \,, \label{Uop} \ee
  we find that $\mathcal{U}(t,t_0)$ obeys
  \be i\frac{d}{dt}\mathcal{U}(t,t_0) = \widetilde{H}^{(H_\chi)}_{I}(t)\, \mathcal{U}(t,t_0) ~~;~~ \mathcal{U}(t_0,t_0)=1\,,
  \label{Uopip} \ee where
  \be \widetilde{H}^{(H_\chi)}_{I}(t) = e^{iH_{\chi} t}\,  {H}_{I}(t)  \, e^{-iH_{\chi} t} =  g \int d^3 x \,\overline{a}(\vx,t)\,\op^{(H_\chi)}(\vx,t)\,,\label{Hchiint}\ee and
  \be \op^{(H_\chi)}(\vx,t)= e^{iH_{\chi} t}\,\op(\vx)\,e^{-iH_{\chi} t}\,,\label{heisOp} \ee is the composite operator in the \emph{Heisenberg} picture in terms of the  Hamiltonian $H_{\chi}$, namely in absence of the coupling to the axion field. The solution of eqn. (\ref{Uopip}) is
  \be \mathcal{U}(t,t_0) = 1- i g \int^t_{t_0}\int \overline{a}(\vx^{\,'},t')\,\op^{(H_\chi)}(\vx^{\,'},t')\,dt'\,d^3 x' + \cdots \label{soluU}  \ee The expectation value of the Schroedinger picture operator $\op(\vx)$ in the non-equilibrium density matrix $\rho_{\chi}(t)$ is
  \be \langle \op(\vx)\rangle (t) = \mathrm{Tr}\Big(  \op(\vx)\,\rho_{\chi}(t)\Big) = \mathrm{Tr}\Big(  \op^{(H_\chi)}(\vx,t)\,\mathcal{U}(t,t_0)\,\rho_{\chi}(t_0)\,\mathcal{U}^{-1}(t,t_0)\Big)\,,\label{expval} \ee where we have used eqns.(\ref{rhochioft},\ref{Uop},\ref{equichi},\ref{heisOp}) and the cyclic property of the trace. Using (\ref{soluU}) up to first order in $g$, and using the cyclic property of the trace, we find
  \be \langle \op(\vx)\rangle (t) = \langle \op(\vx)\rangle (t_0)+  \int d^3 x' \int^t_{t_0} \Xi(\vx-\vx',t-t')\,\overline{a}(\vx^{\,'},t') \, dt' + \cdots \label{expvalO2} \ee where
  \be  \langle \op(\vx)\rangle (t_0) = \mathrm{Tr}\,\op(\vx)\,\rho_{\chi}(t_0)\,.  \label{valto}  \ee The linear response kernel, namely the dynamical susceptibility, is given by
  \be  \Xi(\vx-\vx',t-t')=  -ig \mathrm{Tr}\Big( \Big[\op^{(H_\chi)}(\vx,t),\op^{(H_\chi)}(\vx',t') \Big] \,\rho_{\chi}(t_0) \Big) ~~;~~ t>t' \,, \label{sigma} \ee and $\overline{a}(\vx,t)$ is the solution of the equation of motion for the expectation value of the axion field.  We have used that the equilibrium density matrix is space-translational invariant, and because $[H_{\chi},\rho_{\chi}(t_0)]=0$ it follows that $\Xi$ must be solely a function of $t-t'$, as confirmed by the analysis below.  Assuming that $\rho_{\chi}(t_0)$ is even under parity, it follows that

  \be  \langle \op(\vx)\rangle (t_0) = 0\,. \label{opezero}\ee In section (\ref{sec:discussions}) we discuss the caveats associated with this choice in cosmology.

  Therefore, to leading order in the axion coupling  we find the \emph{induced} non-equilibrium expectation value for $t> t_0$  in linear response
  \be    \langle \op(\vx)\rangle (t) =   \int d^3 x' \int^t_{t_0} \Xi(\vx-\vx',t-t')\,\overline{a}(\vx^{\,'},t')\,dt'\,,\label{finexpval}   \ee with the  dynamical  susceptibility
  $\Xi(\vx-\vx',t-t') $ given by (\ref{sigma}).

It is convenient to write the susceptibility $\Xi$ in terms of a Lehmann (spectral) representation. This is achieved by   writing
\be\op^{(H_\chi)}(\vx,t) = e^{iH_\chi t}\,e^{-i\vec{P}\cdot \vx} \,\op(\vec{0},0) \,e^{-iH_\chi t}\,e^{i\vec{P}\cdot \vx}\,,\label{EPop} \ee and   the density matrix in the   basis   of simultaneous eigenstates of $H_{\chi}$ and the total momentum operator $\vec{P}$, namely  $(H_\chi,\vec{P})\ket{n} = (E_n,\vec{P}_n)\ket{n}$. Introducing the resolution of the identity in this basis $\sum_{m}\ket{m}\bra{m} =1$, and recognizing that $\op^{(H_\chi)}(\vx,t)$ must be a hermitian operator because the axion field is real and the total Hamiltonian is hermitian, we find
 \begin{eqnarray}
\mathrm{Tr}\Big(  \op^{(H_\chi)}(\vx,t)\,\op^{(H_\chi)}({\vx}^{\,'},t')   \,\rho_{\chi}(t_0) \Big) & = &  \frac{1}{\mathrm{Tr}e^{-\beta H_{\chi}}}~
\sum_{m,n}e^{-\beta E_n}
|\langle n| {\cal O}_\chi(\vec{0},0) |m \rangle|^2  \, e^{i(E_n-E_m)(t-t')}\,e^{-i(\vec{P}_n-\vec{P}_m)\cdot(\vx-\vx')}\, \nonumber \\
\mathrm{Tr}\Big(  \op^{(H_\chi)}({\vx}^{\,'},t')\,\op^{(H_\chi)}(\vx,t)   \,\rho_{\chi}(t_0) \Big) & = &  \frac{1}{\mathrm{Tr}e^{-\beta H_{\chi}}}~
\sum_{m,n}e^{-\beta E_n}
|\langle n| {\cal O}_\chi(\vec{0},0) |m \rangle|^2  \, e^{-i(E_n-E_m)(t-t')}\,e^{i(\vec{P}_n-\vec{P}_m)(\vx-\vx')}\,. \nonumber \\ &  & \label{reps}
 \end{eqnarray}  In terms of the spectral functions
\begin{eqnarray}
\rho^>(k_0,\vk) & = &  \frac{(2\pi)^4}{\mathrm{Tr}e^{-\beta H_{\chi}}}~
\sum_{m,n}e^{-\beta E_n}
|\langle n| {\cal O}_\chi(\vec{0},0) |m \rangle|^2  \, \delta(k_0-(E_m-E_n))\,\delta(\vec{k}-(P_m-P_n))\, \label{siggreat} \\
\rho^<(k_0,\vk) & = &  \frac{(2\pi)^4}{\mathrm{Tr}e^{-\beta H_{\chi}}}~
\sum_{m,n} e^{-\beta E_n}
 |\langle n| {\cal O}_\chi(\vec{0},0) |m \rangle|^2  \, \delta(k_0-(E_n-E_m))\,\delta(\vec{k}-(P_n-P_m))\,
 \label{sigless} \,,
\end{eqnarray}
the correlation functions (\ref{reps}) can be written as
\begin{eqnarray}
\mathrm{Tr}\Big(  \op^{(H_\chi)}(\vx,t)\,\op^{(H_\chi)}(\vx^{\,'},t')   \,\rho_{\chi}(t_0) \Big) & = &   \int \frac{d^4k}{(2\pi)^4} \rho^>(k_0,\vk)\, e^{-ik_0(t-t')}\,e^{i\vk\cdot(\vx-\vx')}\, \label{Ogreat} \\
\mathrm{Tr}\Big( \op^{(H_\chi)}(\vx^{\,'},t')\, \op^{(H_\chi)}(\vx,t)   \,\rho_{\chi}(t_0) \Big) & = &   \int \frac{d^4k}{(2\pi)^4} \rho^<(k_0,\vk)\, e^{-ik_0(t-t')}\,e^{i\vk\cdot(\vx-\vx')}\,.  \label{Oless}
 \end{eqnarray}  Upon relabelling
$m \leftrightarrow n$ in the sum in the definition (\ref{sigless}) and recalling that $\mathcal{O}_{\chi}$ is an hermitian operator,
we find the Kubo-Martin-Schwinger relation\cite{kms}
\begin{equation}
\rho^<(k_0,k)  = \rho^>(-k_0,k) = e^{-\beta k_0}
\rho^>(k_0,k)\,. \label{KMS}
\end{equation}
Introducing the spectral density
\be \rho(k_0,k)= \rho^>(k_0,k)-\rho^<(k_0,k)\,,\label{specfun}\ee the relation (\ref{KMS}) implies that
\be \rho(k_0,k) = - \rho(-k_0,k)\,. \label{rhodd} \ee
The dynamical susceptibility is now expressed solely in terms of the spectral density $\rho(k_0,k)$ as
\be  \Xi(\vx-\vx',t-t')=  -ig \,\int\frac{d^4k}{(2\pi)^4}\,\rho(k_0,k)\,e^{-ik_0(t-t')}\,e^{i\vk\cdot(\vx-\vx')}\,.\label{sigfin} \ee

The Lehmann representations (\ref{reps}) and the spectral densities (\ref{siggreat},\ref{sigless}) are \emph{exact} results, valid to all orders in the couplings of the $\chi$ fields to degrees of freedom  within or beyond the standard model expect the axion. Therefore the dynamical susceptibility (\ref{sigfin}) while linear in the coupling $g$ (linear response) is in principle to all orders  in all other couplings.

 In ref.\cite{shuyang} it was found that the expectation value of the axion field obeys the equation of motion
  \be \frac{\partial^2}{\partial t^2} \, \overline{a}(\vx,t)-\nabla^2 \overline{a}(\vx,t)+m^2_{0a}\,\overline{a}(\vx,t)+\int  \int^t_{t_0} \Sigma(\vx-{\vx}^{\,'},t-t')\,\overline{a}({\vx}^{\,'},t')\,d^3x'\,dt' = 0 \,,\label{eqnofmot} \ee where $m^2_{0a}$ is the bare axion mass, and  the retarded self energy $ \Sigma(\vx-{\vx}^{\,'},t-t')$  is given by\cite{shuyang}
  \be \Sigma(\vx-{\vx}^{\,'},t-t') =  -ig^2 \, \mathrm{Tr}\Big( \Big[\op^{(H_\chi)}(\vx,t),\op^{(H_\chi)}(\vx',t') \Big] \,\rho_{\chi}(t_0) \Big)\, \,. \label{selfen} \ee

  Remarkably, the dynamical  susceptibility is simply related to the axion retarded self-energy to leading order in the axion coupling $g$ but to all orders in the couplings of the fields $\chi$ to any other field within or beyond the standard model except the axion\cite{shuyang}, namely
  \be \Sigma(\vx-\vx',t-t') = g\, \Xi(\vx-\vx',t-t'). \label{sigxieq}\ee This is one of the important results of this study, and applies in general for any of the interactions of the form $g \, a(\vx,t)\,\mathcal{O}_{\chi}(\vx,t)$, with important consequences explored below.

The main result of this section is the non-equilibrium induced expectation value of the composite operator $\op$, which in linear response is given by

 \be  \langle \op(\vx)\rangle (t) = -ig \,\int\frac{d^4k}{(2\pi)^4}\,\rho(k_0,k)\, \int  d^3 x' \int^t_{t_0} e^{-ik_0\,(t-t')}\,e^{i\vk\cdot (\vx-\vx^{\,'}) } \,\overline{a}(\vx^{\,'},t')\,dt'\,,\label{finexpval2}   \ee which is obtained by combining equation (\ref{finexpval}) with the  spectral representation (\ref{sigfin}).  This expression can be written in a more illuminating manner by recognizing that $\overline{a}(\vx ,t)$ is the solution of the equation of motion, eqn. (\ref{eqnofmot}), with the   self energy to leading order in the coupling $g$ given by eqn. (\ref{selfen}). Using the relation (\ref{sigxieq}) between the dynamical susceptibility and the self energy, and the equation of motion (\ref{eqnofmot}), it is straightforward to confirm the result (\ref{eqn:exactO}),  which has been obtained as the expectation value of the exact Heisenberg equations of motion,   to leading order in the coupling $g$, namely linear response.

 Note that because $\Sigma \propto g^2$, it follows that
 \begin{equation}
    \frac{\partial^2}{\partial t^2} \, \overline{a}(\vx,t)-\nabla^2 \overline{a}(\vx,t)+m^2_{0a}\,\overline{a}(\vx,t) \propto g^2 \,,
\end{equation} and  the expectation value $\langle \op(\vx)\rangle (t) \propto g$.

\subsection{$\langle \mathcal{O}\rangle $ for misaligned initial conditions.}

  We now consider the cosmologically relevant case of misaligned initial conditions for the axion condensate, where the axion-like fields are produced non-thermally and undergo a damped oscillations. In ref. \cite{shuyang} it is shown that in Minkowski space time the solution of the   equation of motion  (\ref{eqnofmot})  is described by exponentially damped oscillations, in which the frequency and decay rate are spatial-momentum and temperature dependent.   Therefore, a general form of the axion-field amplitude is
\bea
    \overline{a}(\Vec{x},t) & = &  \int \frac{d^3k}{(2\pi)^3}\,e^{i \vk\cdot \vec{x}}\,\overline{a}_k(t) \,,\nonumber \\   \overline{a}_k(t) & = & \Big[ A_k e^{-i \omega_k (t-t_0)} + A_k^* e^{i \omega_k (t-t_0)} \Big]\,e^{-\frac{\Gamma_k}{2}\,(t-t_0)}
    \label{eqn:misaligned-a-bar}
\eea
where $A_k$ is a classical complex amplitude determined by initial conditions, $\omega_k$ is the renormalized frequency with $\omega^2_k+\delta \omega^2_k  = k^2+m^2_{0a} $, $\delta \omega_k$ the renormalization counterterm,  and  $\Gamma_k$ the renormalized decay rate, both depending on momentum and temperature. In appendix (\ref{app:drg}) we provide an alternative method to solve the equation of motion based on multi-time scale analysis yielding the same results.

Taking the spatial Fourier transform of  (\ref{finexpval2}) with the solution(\ref{eqn:misaligned-a-bar}) yields
\begin{equation}
    \langle\mathcal{O}_\chi\rangle_k(t) = -i g \int \frac{dk_0}{2\pi} \int_{t_0}^t dt' \rho(k_0,k) e^{-ik_0 (t-t')} \overline{a}_k(t')\,.
    \label{eqn:Ochik-rho}
\end{equation}
 The time integrals are straightforwardly evaluated with    $\overline{a}_k$  given by eqn. (\ref{eqn:misaligned-a-bar}), yielding
\begin{align}
    \langle\mathcal{O}_\chi\rangle_k(t) =
    & - g \,\Bigg\{ A_k\, e^{-i\Omega_k t}  \,\int_{-\infty}^\infty \frac{dk_0}{2\pi} \frac{\rho(k_0,k)}{k_0 - \Omega_k}
      +A_k^*\, e^{i\Omega^*_k t} \,  \int_{-\infty}^\infty \frac{dk_0}{2\pi} \frac{\rho(k_0,k)}{k_0 + \Omega_k^*} \Bigg\}
    \nonumber \\
    & +g  \,\Bigg\{ A_k\, e^{-i\Omega_k t_0}  \,\int_{-\infty}^\infty \frac{dk_0}{2\pi} \frac{\rho(k_0,k)}{k_0 - \Omega_k}\,e^{-ik_0(t-t_0)}\,
      +A_k^*\, e^{i\Omega^*_k t_0} \,  \int_{-\infty}^\infty \frac{dk_0}{2\pi} \frac{\rho(k_0,k)}{k_0 + \Omega_k^*}\,e^{-ik_0(t-t_0)} \Bigg\}\,. \label{Okt}
\end{align} where $\Omega_k = \omega_k - i \Gamma_k/2$.

The second line in eqn. (\ref{Okt})  features  oscillatory integrals that represent transient processes due to the sudden switch-on at time $t=t_0$,  and vanish fast for    $t-t_0\gg m_a$, therefore this contribution will be neglected\footnote{Similar contributions are neglected in obtaining the solution (\ref{eqn:misaligned-a-bar}), see appendix (\ref{app:drg}) and ref.\cite{shuyang} for details.}.

 In the first line, we use the narrow width approximation $\Gamma_k \rightarrow 0$ and the identity $1/(x\pm i0^+) = \mathcal{P}(1/x) \mp i \pi \delta(x)$ and obtain to leading order in g,

\begin{equation}
    \langle\mathcal{O}_\chi\rangle_k (t) = \frac{1}{g}\,\Bigg[ \Sigma_R(k,\omega_k)\, \overline{a}_k(t) +  {\Gamma_k} \,\dot{\overline{a}}_k(t)\Bigg] \,, \label{finOx}
\end{equation}
where\cite{shuyang} (see also appendix (\ref{app:drg}))
\begin{equation}
    \Sigma_R(k,\omega_k) = g^2 \int_{-\infty}^\infty \frac{dk_0}{2\pi}\mathcal{P} \Bigg[  \frac{\rho(k_0,k)}{\omega_k - k_0}\Bigg]~~;~~ \frac{g^2}{2}\, \rho(\omega_k,k) = \omega_k\,\Gamma_k \,.
    \label{eqn:self-energy}
\end{equation}

This result is in complete agreement with the result (\ref{eqn:exactO}) as can be seen as follows: taking the spatial Fourier transform of eqn. (\ref{eqn:exactO}), using the expression for $\overline{a}_k(t)$ given by eqn. (\ref{eqn:misaligned-a-bar}), and neglecting terms of order $\Gamma^2_k \propto g^4$ we find
\be   \langle\mathcal{O}_\chi\rangle_k (t) = \frac{1}{g}\Bigg[ \Big(\omega^2_k-(k^2+m^2_{0a})\Big)\,\overline{a}_k(t)+\Gamma_k \,\dot{\overline{a}}_k(t)\Bigg]\,,\label{aveOk}  \ee the exact frequency $\omega_k$ corresponds to the real part of the pole in the propagator of the axion field, namely it is the solution of the equation\cite{shuyang}
\be \omega^2_k=(k^2+m^2_{0a})+\Sigma_R(\omega_k,k)\,, \label{poleeqn}\ee from which equation (\ref{finOx}) follows, thereby explicitly confirming the equivalence of the results (\ref{eqn:exactO}, \ref{finOx}) in linear response.

 \vspace{1mm}

 \section{Chern Simons condensate:}\label{sec:CS}

\begin{figure}[h!]
\includegraphics[height=2.5in,width=2.5in,keepaspectratio=true]{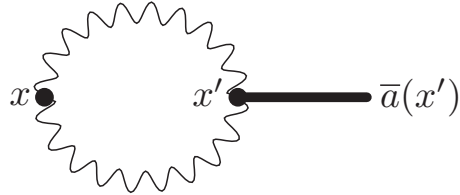}
\caption{$\langle \vec{E}\cdot \vec{B}\rangle (x)$ in linear response (\ref{csexpval}) for the case of free photons. The wavy lines correspond to the retarded correlation function (\ref{CSsus}), the heavy solid line to the misaligned axion condensate $\overline{a}(x')$.}
\label{fig:ebcon}
\end{figure}

 The results above are valid for any generic pseudoscalar composite coupled to the axion field as in the Lagrangian density (\ref{lag}). We now focus specifically on the case of the axion field coupled to photons via the Chern-Simons term
   \be \op(\vx,t) = \vec{E}(\vx,t) \cdot \vec{B}(\vx,t)= -\frac{1}{4} F_{\mu \nu}\widetilde{F}^{\mu \nu}~~;~~\widetilde{F}^{\mu \nu} = \frac{1}{2}\, \varepsilon^{\mu\nu\alpha\beta}\,F_{\alpha\beta}\,, \label{CSdens}  \ee this   pseudoscalar density    is a total surface term, since
   \be F_{\mu \nu}\widetilde{F}^{\mu \nu} \propto \partial_{\mu}\Big(\varepsilon^{\mu\nu\alpha\beta} A_{\nu} \partial_{\alpha} A_{\beta}\Big) \,.\label{surfterm}\ee
   In this case
   \be \langle \vec{E}\cdot \vec{B}\rangle (\vx,t) =   \int d^3 x' \int^t_{t_0} \Xi(\vx-\vx',t-t')\,\overline{a}(\vx^{\,'},t')\,dt'\,,\label{csexpval}  \ee where  the susceptibility is the retarded commutator of the Chern-Simons density
   \be \Xi(\vx-\vx',t-t')=  -ig \mathrm{Tr}\Big( \Big[(\vec{E}\cdot\vec{B})^{(H_{em})}(\vx,t),(\vec{E}\cdot\vec{B})^{(H_{em})}({\vx}^{\,'},t') \Big]\,\rho_{\chi}(t_0) \Big)\,\Theta(t-t') \,,\label{CSsus}  \ee to which we refer as the \emph{Chern-Simons susceptibility} in analogy with response functions in many body physics. The superscript $(H_{em})$ in the operators of the Chern-Simons density refer to the   Heisenberg fields in absence of their coupling to the axion, namely to all orders in the electromagnetic interaction with degrees of freedom within or beyond the standard model except for the axion field. The Feynman diagram describing the induced condensate (\ref{csexpval}) in the case of free photons is shown in fig. (\ref{fig:ebcon}).

   We now obtain the Chern-Simons susceptibility $\Xi$ (\ref{CSsus}) by considering free electromagnetic fields. This approximation is valid within the cosmological setting after recombination for the following reasons: when the temperature ($\simeq \mathrm{eV}$) is much smaller than the electron mass  the lepton contribution to the renormalized photon self-energy is perturbatively small and  thermally suppressed, therefore there is no (gauge invariant) thermal mass for the photon\cite{htl,lebellac}. Furthermore the free electron density $n$  vanishes rapidly during recombination, therefore the plasma frequency $\Omega_{pl}=  \sqrt{4\pi e^2 n/m}$ is vanishingly small and the photon bath is described by blackbody radiation, namely free thermal photons, as evidenced by the (nearly) blackbody spectrum of the cosmic microwave background.

       Under these conditions, the spectral density (\ref{specfun}) has been obtained in refs.\cite{shuyang}, and is given by
 \bea
  &&  \rho(k_0,\vec{k})
     =   \frac{(K^2)^2}{32\pi}\,\Bigg\{\Bigg(1 + \frac{2}{\beta k}\,\ln\Bigg[\frac{1-e^{-\beta \omega^I_+}}{1-e^{-\beta \omega^I_-}} \Bigg]\Bigg)\,\Theta(K^2)  + \frac{2}{\beta k}\, \ln\Bigg[\frac{1-e^{-\beta \omega^{II}_+}}{1-e^{-\beta \omega^{II}_-}} \Bigg]\,\Theta(-K^2) \Bigg\}\, \mathrm{sign}(k_0)\,, \nonumber  \\  && K^2=k^2_0-k^2 ~~;~~   \omega_\pm^{(I)}   =   \frac{|k_0| \pm k}{2}~~;~~
    {\omega}_\pm^{(II)} = \frac{k \pm |k_0|}{2}\,. \label{rhofiI} \eea The terms with $\Theta(k^2_0-k^2)$ arise from the processes $a \leftrightarrow 2 \gamma$, namely emission and absorption of photons with the reverse or recombination process $2\gamma \rightarrow a$  being a consequence of the radiation bath, these processes  feature support on the axion mass shell for massive axions. The contribution proportional to $\Theta(k^2-k^2_0)$ only features support below the light cone, it describes off-shell processes $\gamma a \leftrightarrow \gamma$ and vanishes in the $k\rightarrow 0$ limit.

Motivated by the cosmological case we now consider a homogeneous misaligned axion condensate depending solely on time  by setting
\be A_k = (2\pi)^3\,\delta^3(\vk)\,a_0   \,, \label{homoaxion} \ee   namely
\be \overline{a}(t)=  e^{-\frac{\Gamma}{2}\,t}\,\big( a_0 \, e^{-im_at}+ a^*_0 \, e^{im_at}\big) \,, \label{misosci}\ee from which it follows that the induced Chern-Simons condensate is also homogeneous and from the result (\ref{eqn:Ochik-rho}) it is given by
\be \langle \vec{E}\cdot \vec{B} \rangle (t)=  -ig \,\int_{-\infty}^{\infty}\frac{dk_0}{(2\pi)}\,\rho(k_0,0)\,e^{-ik_0\,t}    \int^t_{t_0} e^{ik_0\,t'}   \,\overline{a}(t')\,dt'\,.\label{CSC} \ee  From eqn. (\ref{rhofiI}) we find
\be \rho(k_0,0) =  \frac{ k^4_0}{32\pi}\, \Bigg (1 +2\, n \Big(\frac{k_0}{2}\Big) \Bigg)\,\mathrm{sign}(k_0)~~;~~ n(\omega)= \frac{1}{e^{\beta \omega}-1}\,. \label{rhoze}  \ee

Before we study the response to an oscillating coherent misaligned $\overline{a}(t)$, it is illuminating to consider the case wherein such expectation value has relaxed to a time independent equilibrium minimum $\overline{a}_0$ at a time $t_0$ and remains constant for  $t>t_0$. Such situation emerges from a damped oscillatory expectation value around a  minimum away from the origin, if, for example   the axion potential features such a minimum. Setting $\overline{a}(t') \rightarrow \overline{a}_0$ for $t> t_0$,  the time integral in (\ref{CSC})  becomes $\propto \sin\big(k_0(t-t_0)/2\big)/k_0$ yielding a non-vanishing Chern-Simons condensate. However  as the time interval $t-t_0\rightarrow \infty$ the time integral $\rightarrow  2\pi\,\delta(k_0)$ and the induced Chern-Simons condensate vanishes. This is consistent with the fact that the Chern-Simons density $\vec{E}\cdot\vec{B}$ is a total surface term (\ref{surfterm})\cite{carroll}, and its space-time integral vanishes in the infinite time and volume limit. However, the non-equilibrium result within \emph{finite   time-like hypersurfaces} is non-vanishing, and as the time integral in (\ref{CSC}) makes explicit, the induced condensate is proportional to the difference of a function evaluated at the two hypersurfaces at times $t$ and $t_0$, when  the spatial volume has been taken to infinity. Furthermore, $\overline{a}(t)$ is the dynamical axion condensate which is a solution of the equation of motion (\ref{eqnofmot}) for $k=0$, and we note that the only space-time constant solution of the equation of motion (\ref{eqnofmot}) is $\overline{a} =0$, which obviously yields a vanishing Chern-Simons condensate as suggested in eqn. (\ref{eqn:exactO}).

Within the cosmological setting, the initial time $t_0$ is approximately the time of the last scattering surface because we are considering free photons in thermal equilibrium in the intermediate state, and $t$ is of the order of the Hubble time today so that $t \gg t_0$ and $t-t_0 \gg 1/m_a, 1/T$.

Let us now consider a dynamical homogeneous   solution of the equation of motion (\ref{eqnofmot}) given by eqn. (\ref{misosci})
where $m_a$ is the renormalized axion mass\cite{shuyang}, and   $a_0$ is a classical complex amplitude determined by initial conditions (see appendix (\ref{app:drg})).   From the general result (\ref{aveOk}) for $k=0$    with $\omega_{k=0}\equiv m_a$, and keeping consistently terms up to $\mathcal{O}(g^2)$ we find,
\be  \langle \vec{E}\cdot \vec{B} \rangle (t) =  \frac{1}{g}  \, \Bigg[ (m^2_a-m^2_{0a})\,\overline{a}(t)+  {\Gamma} \,\dot{\overline{a}}(t) \Bigg] \,. \label{exeb} \ee
Furthermore, the relations (\ref{poleeqn},\ref{eqn:self-energy}) yield
\be m^2_a - m^2_{0a} = g^2 \int_{-\infty}^{\infty} \mathcal{P}\Bigg[\frac{\rho(k_0,0)}{m_a-k_0} \Bigg] \,\frac{dk_0}{2\pi} = \Sigma_R(0,m_a) \,,\label{renmas}\ee
 and\cite{shuyang}
\be \Gamma = g^2\,\frac{\rho(m_a,0)}{2\,m_a}= \frac{g^2\,m^3_a}{64\pi} \, \Bigg(1 +  {2}\,n\Big(\frac{m_a}{2}\Big)\Bigg)  \,,\label{gamma}\ee is the axion decay rate $a\rightarrow 2 \gamma$\cite{shuyang} (see appendix (\ref{app:drg})).

 This is one of the main results of this study. For the case of an ultralight axion witn $m_a \lesssim \mu \mathrm{eV}$ and even for temperatures of the order of the (CMB) temperature today  $10^{-4}\,\mathrm{eV}$, it follows that $T \gg m_a$. Therefore using the spectral density (\ref{rhoze}) in this high temperature limit we find\cite{shuyang}
 \be m^2_a - m^2_{0a} = -\frac{g^2\,\Lambda^4}{128\pi^2}\,\mathcal{F}_{0}\Big[\frac{m^2_a}{\Lambda^2}\Big]- \frac{g^2\,\pi^2\,T^4}{15}\,\mathcal{F}_{T}\Big[\frac{m^2_a}{T^2}\Big]\,,\label{deltam2} \ee where the first term is the zero temperature contribution, for which we carried out the integral in $k_0$  in (\ref{renmas}) with an ultraviolet   cutoff $\Lambda\gg m_a, T$,   delimiting the regime of validity of the effective field theory, and the second term is the finite temperature contribution, with
\bea & & \mathcal{F}_{0}\Big[\frac{m^2_a}{\Lambda^2}\Big] = 1+  4\,\frac{m^2_a}{\Lambda^2} +\frac{m^4_a}{\Lambda^4}\ln\Big( \frac{\Lambda^2}{m^2_a}\Big) + \cdots \label{vacon} \\ & & \mathcal{F}_{T}\Big[\frac{m^2_a}{T^2}\Big] =1 + \frac{15\,m^2_a}{24\,\pi^2\,T^2}- \frac{15\,m^4_a}{32\,\pi^2\,T^4}\,\ln\Big[\frac{T}{m_a}\Big]+ \cdots  \label{finiteT} \eea  The dots stand for higher orders in $m_a/\Lambda$ and $m_a/T$ respectively, and to leading order in the high temperature limit we find
\be \Gamma = \frac{g^2\,m^2_a\,T}{16\,\pi}\,. \label{hiTgam} \ee

The ultraviolet divergence of the Chern-Simons condensate resulting from the first term in (\ref{deltam2}) is not unexpected. The Chern-Simons density is an operator of mass dimension four, just like $\vec{E}^2$ and $\vec{B}^2$, however, unlike these operators whose vacuum expectation values yield the zero point energy, the expectation value of $\vec{E}\cdot\vec{B}$ vanishes if the state is invariant under parity. In other words,  the expectation value of the Chern-Simons density is protected from ultraviolet divergences by parity. However, an expectation value of the pseudoscalar axion   breaks parity, therefore, in presence of this parity-symmetry breaking term  the ultraviolet divergence of $\vec{E}\cdot\vec{B}$ becomes explicit, and as exhibited by eqn.
(\ref{exeb}) is proportional to the symmetry breaking term in linear response. We choose to renormalize the Chern-Simons condensate by subtracting the $T=0$ (vacuum) contribution,  subtle aspects of  renormalization   are discussed in more detail in section (\ref{sec:discussions}). Keeping the leading order term in the high temperature expansion,  the renormalized expectation value is given by
\be    \langle \vec{E}\cdot \vec{B} \rangle^{(R)} (t) =  -\frac{g\,\pi^2\,T^4}{15}    \,\overline{a}(t)+ \frac{g \,m^2_a\,T}{16\,\pi}\,\dot{\overline{a}}(t) + \mathcal{O}(m^2_a/T^2) \,, \label{cscondren} \ee with $\overline{a}(t)$ the (spatially homogeneous) solution of the equation of motion (\ref{eqnofmot}).  This is another of  the main results of this study.

We can obtain an estimate of the energy density stored in the Chern-Simons condensate  as compared to that in the cosmic microwave background today, $\rho_{0\gamma}= \pi^2 T^4_{0\gamma}/15$,  by using the following estimates:
\be m^2_a \, a^2 \simeq \rho_{0DM} = \rho_{0c}\,\Omega_{DM}~~;~~ g = \frac{\mathcal{C}}{f_a}\,, \label{estis}\ee where
$\mathcal{C}<1$ is a dimensionless constant, $f_a$ the axion decay constant, and using the values, $\Omega_{DM} \simeq 0.23; h \simeq 0.7$ and the temperature of the cosmic microwave background today $T_{0\gamma} = 2.37\times 10^{-4}\,\mathrm{eV}$, we find
\be \Big| \frac{\langle \vec{E}\cdot \vec{B} \rangle^{(R)} (t_0)}{\rho_{0\gamma}} \Big| \simeq \mathcal{C}\,\Big(\frac{10^{10}\,\mathrm{GeV}}{ f_a}\Big)\,\Big(\frac{\mu\,eV}{m_a}\Big)\,\Bigg[1- 2.2\times 10^{-9}\,\Big(\frac{m_a}{\mu\,eV} \Big)^3\Bigg] \times 3\times 10^{-19}\,. \label{enerdens} \ee

This analysis suggests that low mass axions with $m_a \ll \mu eV$  yield larger contributions to the energy density stored in the Chern Simons condensate, perhaps leading to an observational avenue.

\vspace{1mm}

\section{Probing the Chern-Simons condensate with emergent axion quasiparticles:}\label{sec:emeraxion}

The analysis above unambiguously implies that a macroscopic  axion condensate will induce a macroscopic Chern-Simons condensate, leading to the question of what are the observational consequences of such topological condensate. While it is possible that a cosmological imprint of this condensate may be observable in the polarization signal of the (CMB) in a manner yet to be understood and studied further, here we \emph{suggest} that the emergent axion-like quasiparticles in topological insulators\cite{wang,narang,gooth}, Weyl semimetals\cite{yu}, multilayered metamaterials\cite{wilczek}, or magnetoelectric insulators\cite{binek}  may be harnessed to probe both the cosmological axion \emph{and} the Chern-Simons condensate. In these materials, an axion-like collective quasiparticle excitation $\Theta(\vx,t)$ couples to the electromagnetic fields\cite{wilczekaxion,wang,narang,gooth,yu,binek} with
\be \mathcal{L}_{\Theta} = \alpha  \Theta(\vx,t)\,\vec{E}(\vx,t)\cdot \vec{B}(\vx,t) \,,\label{tetalag} \ee with $\alpha$ the electromagnetic fine structure constant. In multilayered metamaterials, instead of $\alpha$  the effective coupling can be tuned making these platforms more flexible\cite{wilczek}.
This coupling brings \emph{two} important consequences, both relevant to probing cosmological axions:
\begin{itemize}
\item \textbf{i:)} the emergent axion-like quasiparticles described by the effective field $\Theta(\vx,t)$ \emph{mix} with the cosmological axion $a(\vx,t)$ via a common two-photon intermediate state. This is depicted in fig. (\ref{fig:mixing}) by the photon loop connecting the external fields $\Theta(\vx,t)$ and $a(\vx,t)$ resulting in off-diagonal components of the propagators in the material. An important aspect of this mixing is that
the off-diagonal matrix elements of the propagator is of order $g\,\alpha$.  This aspect combined with     coherence of the axion field in the form of a macroscopic condensate may yield observational effects at leading order in $g$, which results in an enhancement in detection efficiency over other possible processes such as   ``axion shining through walls'' with a transition probability of order $g^4$.

        \begin{figure}[h!]
\includegraphics[height=2.5in,width=2.5in,keepaspectratio=true]{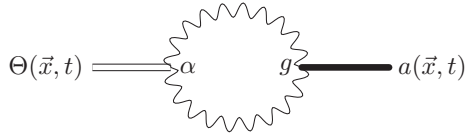}
\caption{Mixing between the emergent axion field $\Theta(\vx,t)$ and the cosmological axion field $a(\vx,t)$ via the Chern-Simons correlation function. The wavy lines correspond to the   correlation function (\ref{CSsus}), the heavy solid line to the    axion field $ {a}(x')$, and the double solid line to the
emergent axion quasiparticle field $\Theta(\vx,t)$.}
\label{fig:mixing}
\end{figure}

\item \textbf{ii:)} As discussed above, in presence of a misaligned (cosmological) axion condensate, $a(\vx,t) = \overline{a}(\vx,t)+\widetilde{a}(\vx,t)$ where $\overline{a}(\vx,t)$ is a classical field describing a macroscopic condensate. Replacing $a(\vx,t) \rightarrow \overline{a}(\vx,t)$ in the external leg of the mixed propagator in fig. (\ref{fig:mixing}), the photon loop and the external c-number external leg $\overline{a}(\vx,t)$ yield the Chern-Simons condensate (see fig. (\ref{fig:ebcon})) $\langle \vec{E}\cdot\vec{B}\rangle (\vx,t)$ which acts as an external time dependent c-number source term, namely $\alpha \Theta(\vx,t)  \langle \vec{E}\cdot\vec{B}\rangle (\vx,t) \rightarrow \alpha\,\Theta(\vx,t)\,h(\vx,t)$, as depicted in fig. (\ref{fig:source}).

        \begin{figure}[h!]
\includegraphics[height=3.2in,width=3.2in,keepaspectratio=true]{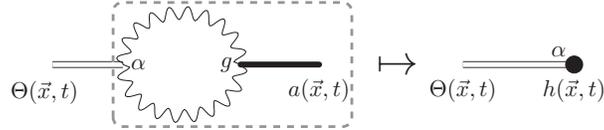}
\caption{Replacing the axion field by the misaligned axion condensate $\overline{a}(\vx,t)$ the photon loop with $\overline{a}$ as the external leg is identified with the Chern-Simons condensate $\langle \vec{E}\cdot\vec{B}\rangle$ (see fig.(\ref{fig:ebcon})), which acts as an external source $h(\vx,t)=\langle\vec{E}\cdot\vec{B}\rangle$ linearly coupled to the quasiparticle axion field $\Theta(\vx,t)$. }
\label{fig:source}
\end{figure}

\end{itemize}

Such external source term, linearly coupled to $\Theta(\vx,t)$ results in an effective external time dependent driving term displacing the quasiparticle field off-equilibrium with an oscillatory behavior corresponding to the time dependence of the \emph{cosmological} axion. For a homogeneous axion condensate, this driving term induces an oscillatory macroscopic condensate of the emergent axion-like field in the material, namely a coherent state of the quasiparticle degrees of freedom, which in principle could be measured along the lines of the
experimental setup in ref.\cite{binek}, and perhaps with enhanced tunability of the coupling in the case of multilayered metamaterials\cite{wilczek}, thereby directly probing the Chern-Simons condensate and, indirectly,  the cosmological axion condensate. However a recent analysis of the signal to noise ratio in multiferroics suggests that the coupling between axion dark matter and ferroic orders in multiferroics\cite{bala} may not yield an observable signal of dark matter axions. However, the possibility of harnessing other materials for detection, in particular via the coupling to the Chern Simons condensate remains to be explored.

\vspace{1mm}

\section{Discussions:}\label{sec:discussions}

\textbf{Renormalization of the Chern-Simons condensate:} As discussed above, for a non-vanishing dynamical expectation value of the axion field, the induced Chern-Simons condensate features ultraviolet divergences and it must be renormalized. The first term in (\ref{exeb}) and the function $\mathcal{F}_0$ given by (\ref{vacon}) are obtained by imposing an ultraviolet cutoff $\Lambda$ in the $k_0$ integral of the spectral density in eqn. (\ref{renmas}). This cutoff is interpreted as the scale below which the effective field theory described by the local Lagrangian density (\ref{lag}) is valid. Even when this scale is finite, the result (\ref{exeb}) implies a strong sensitivity to this scale. There does not seem to be an obvious manner to renormalize the Chern-Simons condensate since it depends explicitly on time through  the dynamical expectation value of the axion field. Therefore,  we proceed to renormalize it simply by subtracting the vacuum contribution, yielding  the leading order renormalized condensate in the high temperature limit $T\gg m_a$    given by the result (\ref{cscondren}).
  The vacuum subtraction is motivated by the subtraction of the zero point contributions to $\langle \vec{E}^2 \rangle$ and $\langle \vec{B}^2 \rangle$, namely the subtraction of the zero point energy, since these operators are also of mass dimension four and feature the same type of ultraviolet divergences $\propto \Lambda^4$ as the Chern-Simons condensate. However, unlike the vacuum subtraction for $\langle \vec{E}^2 \rangle,\langle \vec{B}^2 \rangle$ the subtraction for  $ \langle \vec{E}\cdot \vec{B} \rangle $ is proportional to  the misaligned axion condensate which depends explicitly on time. It remains to be explored further if there is a suitable and more rigorous renormalization scheme for $\langle \vec{E}\cdot \vec{B}\rangle(t)$, beyond a vacuum subtraction or subtracting solely the ultraviolet sensitive
  terms.

\vspace{1mm}

\textbf{Cosmological caveats:}    Our ultimate objective is to understand the cosmological implications of the non-equilibrium dynamics of axions (or axion-like particles) in cosmology. To this aim, the results obtained above in Minkowski space time serve as a prelude, and a ``proof of principle'' of the application of the concepts behind  linear response to extract the induced parity violating Chern-Simons condensate. There are several obvious differences between the dynamics in Minkowski space time, and in cosmology: Hubble  expansion modifies the time evolution of the axion condensate including a damping term in the equation of motion proportional to the Hubble rate of expansion,  axion decay into photons, or other processes that lead to damping of the condensate will also add to the damping dynamics but through a self-energy correction that must be obtained from field  quantization in the expanding cosmology. However, by dimensional analysis,  linear response and   under the assumption that the axion condensate undergoes damped oscillations, the general form (\ref{exeb}) qualitatively describes the Chern-Simons condensate , albeit with different functional form of $\Gamma$ and the function $\mathcal{F}_T$   in eqn. (\ref{finiteT}) since the high temperature behavior of $m^2_a-m^2_{0a} \propto g^2 T^4$ on dimensional grounds, and both must include the effect of Hubble  expansion.

These aspects notwithstanding, the results in Minkowski space time indicate that the \emph{qualitative} aspects and main conclusion, namely a dynamical misaligned coherent axion condensate will \emph{induce} a macroscopic condensate of the composite operator(s) coupled to the axion as in eqn. (\ref{lag}),  will remain. Therefore, the calculation in Minkowski space-time   with the approximation of free photons in the Chern-Simons susceptibility provides a ``proof of principle'' of the main concepts and the qualitative form of the condensate.

Furthermore, the potential observational consequences of such condensate in topological or metamaterials as discussed in the previous section are reliably described by the Chern-Simons susceptibility calculated with free photons as such possible experiments would be carried out today when the radiation bath to which the cosmological axion is coupled is the cosmic microwave background.

\vspace{1mm}

\textbf{Neutral pion condensate from misaligned axions:}
 The axion is a quasi Nambu-Goldstone boson, and as such it couples \emph{directly} to other matter fields via a derivative coupling to a pseudovector current. However, the axion couples \emph{indirectly} to the neutral pion via an intermediate state of two photons as can be understood with the following argument. The neutral pion decays into two photons, with an effective coupling of the form $\frac{\alpha}{8\pi\,f_\pi}\,\pi^0\,F_{\mu\nu}\widetilde{F}^{\mu \nu} \simeq \pi^0 \vec{E}\cdot \vec{B}$ as a consequence of the chiral anomaly,  with $\alpha/\pi f_\pi \simeq 0.025\,\mathrm{GeV}^{-1}$, with $\alpha$ the fine structure constant and $f_\pi$ the pion decay constant.   This implies the process $\pi^0   \leftrightarrow 2\gamma  \leftrightarrow a$,
  described by a Feynman diagram similar to that in fig. (\ref{fig:source}) but replacing $\Theta \rightarrow \pi^0$. This process entails that  the axion and the neutral pion can mix via a common intermediate state of two photons,  this is an off-diagonal self-energy diagram that is completely determined by the Chern-Simons dynamical susceptibility (\ref{CSsus}), therefore we expect this ``mixing'' to be of order $g\,\alpha/f_\pi \times T^4$. We are currently exploring this phenomenon.

\vspace{1mm}

\textbf{Non-abelian Chern-Simons condensate:} An axion coupling of the form $g\,a(x)\, G_{\mu \nu,b}\,\widetilde{G}^{\mu\nu,b}$ where $G_{\mu \nu,b}$ is the gluon gauge field strength tensor and $ \widetilde{G}^{\mu\nu,b}$ its dual, would   yield a non-abelian Chern-Simons condensate $\langle G_{\mu \nu,b}\,\widetilde{G}^{\mu\nu,b}   \rangle(t) $ in the same way as that for the abelian gauge theory. In this case the dynamical susceptibility is the retarded commutator (\ref{sigma}) but with
$\mathcal{O}_{\chi} = G_{\mu \nu,b}\,\widetilde{G}^{\mu\nu,b}$. While in principle a calculation similar to that of  the abelian case yields the dynamical susceptibility (and the quark-gluon contribution to the axion self energy) there are several important differences with the abelian case that would lead to daunting technical aspects. Not only the non-abelian nature of the gauge field introduces new vertices, but also the fact that below the QCD temperature gluons and quarks are confined to mesons and baryons involving non-perturbative physics. Furthermore for temperatures above the QCD scale, which is larger than the masses of all but the top quark, these degrees of freedom are ultrarelativistic and yield self-energy corrections to the gluon propagators in the form of hard thermal loops both from gluons and quarks which cannot be neglected\cite{htl,lebellac}. On dimensional grounds ($ G_{\mu \nu,b}\,\widetilde{G}^{\mu\nu,b} $ has mass dimension four) we expect ultraviolet divergences, or rather sensitivity to a cutoff $\Lambda$ delimiting the validity of the effective field theory description similar to the abelian case. Furthermore,   we also expect that the non-abelian Chern-Simons condensate will be proportional to $\overline{a}(t),\dot{\overline{a}}(t)$, hence an ambiguity in the renormalization of this condensate should arise in much the same way as for the abelian case. From the general results (\ref{eqn:exactO},\ref{misosci} ) and just on dimensional grounds we expect that after subtracting the vacuum term altogether, the high temperature limit of the finite temperature contribution is of the form
\be \langle G_{\mu \nu,b}\,\widetilde{G}^{\mu\nu,b}   \rangle(t)  =  g\,C_{G}\,T^4\,\overline{a}(t)+ \frac{D_{G}}{g} \,\Gamma(T)\,\dot{\overline{a}}(t) \,, \label{nacs}\ee where  $\Gamma(T)$ is the relaxation rate of the misaligned axion condensate, and  $C_G,D_G$ will be functions of the ratios of the various scales, such as the axion mass and quark masses to the temperature. Notwithstanding these quantitative and technical aspects, the general results (\ref{eqn:exactO},\ref{misosci})  imply  that a misaligned axion will induce a condensate of the non-abelian Chern-Simons density.

 \section{Conclusions}\label{sec:conclusions}

The non-equilibrium dynamics of axions is of an  interdisciplinary interest   both in cosmology, as a possible candidate for dark matter and or dark energy as well as in condensed matter physics where axion-like   excitations may emerge as collective quasiparticles in parity violating topological insulators,  density waves in Weyl semimetals or in metamaterials. In this article we study hitherto unexplored non-equilibrium aspects of axions in Minkowski space time as a stepping stone towards a more comprehensive treatment in cosmology with potential observable consequences harnessing emergent axionic quasiparticles in condensed matter systems.

 The main objective   is to study  how a misaligned coherent axion condensate   induces a macroscopic condensate of  a composite pseudoscalar operator $\mathcal{O}$ coupled to the axion as $g\,a(\vx,t)\,\mathcal{O}(\vx,t)$. To this aim we  implement  the method of linear response ubiquitous in  many body physics. We obtain the macroscopic condensate of such pseudoscalar operator to leading order in the coupling $g$ but in principle to all orders of the couplings of the degrees of freedom described by $\mathcal{O}$ to any other degree of freedom within or beyond the standard model, except the axion.

 We focused in particular on a macroscopic condensate of the   Chern-Simons density $\vec{E}\cdot\vec{B}$ in axion electrodynamics and introduced the   dynamical susceptibility, which is a response function that relates the induced Chern-Simons condensate to the axion condensate. This susceptibility is simply related to the axion self energy a relation that holds to all orders in the couplings of photons to other degrees of freedom within or beyond the standard model except the axion. The induced Chern-Simons condensate is strongly sensitive to the cutoff scale of the effective field theory and requires   subtle renormalization. Subtracting the vacuum contribution and in the high temperature limit $T\gg m_a$ we find
 \be \langle \vec{E}\cdot\vec{B}\rangle(t) =  -\frac{g\,\pi^2\,T^4}{15}    \,\overline{a}(t)+ \frac{g \,m^2_a\,T}{16\,\pi}\,\dot{\overline{a}}(t) + \mathcal{O}(m^2_a/T^2)\ee  with $\overline{a}(t)$ the   dynamical homogeneous axion condensate solution of the equations of motion including the self-energy.

We have argued that the Chern-Simons condensate can be probed by harnessing axion-like collective quasiparticles in condensed matter systems such as topological insulators, Weyl semimetals,   magnetoelectric insulators, or multilayered metamaterials which provide realizations of axion electrodynamics. A homogeneous cosmological axion condensate induces a macroscopic topological Chern-Simons condensate that acts like an external source driving a non-equilibrium macroscopic emergent axion-like condensate in these systems, which oscillates with the same frequency as the cosmological axion. Furthermore, we also conjectured that a misaligned axion condensate also induces a non-abelian Chern-Simons condensate and mixes with neutral pion degrees of freedom thereby inducing a neutral pion condensate and chiral symmetry breaking during or after the QCD phase transition.

The next stage of the study will include cosmological expansion which requires extending the linear response formulation to the realm of an Friedmann-Robertson-Walker cosmology, we expect to report on these studies in a future article.

\acknowledgements
  The authors gratefully acknowledge  support from the U.S. National Science Foundation through grant   NSF 2111743.

\appendix

\section{Solution of (\ref{eqnofmot}) via multi-time scale analysis.}\label{app:drg}
Let us consider the spatial Fourier transform of the eqn. of motion (\ref{eqnofmot}), which we write as
\be \ddot{\overline{a}}_k(t)+[\omega^2_k  +\delta \omega_k^2]\,\overline{a}_k(t)+\int^t_{t_0}  \Sigma(k,t-t')\,\overline{a}_k(t')\,dt' = 0 \,,\label{keqn} \ee where anticipating renormalization, $\omega^2_k $ is the \emph{finite temperature renormalized frequency} describing the in-medium dispersion relation and the counterterm $\delta \omega_k^2$ accounts for  renormalization and is defined so that $k^2+m^2_{0a} = \omega^2_k+\delta \omega_k^2$, and
\be \Sigma(k,t-t')= -i g^2\,\int \frac{dk_0}{2\pi}\,\rho(k_0,k)\, e^{-ik_0(t-t')} \,.\label{kself}\ee   Since $\Sigma \propto g^2$,   we expect that $\delta \omega_k^2  \propto g^2$.  Let us write the solution of (\ref{keqn}) as
\be  \overline{a}_{\vk}(t) = A_{\vk}(t)\,e^{-i\omega_k t}+ A^*_{\vk}(t)\,e^{i\omega_k t}\,, \label{slowalfa} \ee where $e^{\mp i \omega_k t}$ are \emph{fast varying} and $A_{\vk}(t)$ is a slowly varying amplitude, so that $\dot{A}_{\vk}(t), \ddot{A}_{\vk}(t)  \propto g^2,      \cdots$. The equation for the slowly varying   amplitudes $A_{\vk}(t),A^*_{\vk}$ become
\bea & & e^{-i\omega_k t}\Bigg[\ddot{A}_{\vk}-2i\omega_k\,\dot{A}_{\vk} +\delta \omega_k^2 \, A_{\vk}+\int^t_{t_0}  \Sigma(k,t-t')\,e^{i\omega_k(t-t')}\,A_{\vk}(t')\,dt' \Bigg] \nonumber \\
&  + & e^{i\omega_k t}\Bigg[\ddot{A}^*_{\vk}+2i\omega_k\,\dot{A}^*_{\vk} + \delta \omega_k^2 \, A^*_{\vk}+\int^t_{t_0}  \Sigma(k,t-t')\,e^{-i\omega_k(t-t')}\,A^*_{\vk}(t')\,dt' \Bigg] = 0 \,. \label{alfaeqns}\eea Since the fast varying phases are independent and the brackets are of $\mathcal{O}(g^2)$ or higher, we request that each bracket vanishes individually, yielding
 \bea & &  \ddot{A}_{\vk}-2i\omega_k\,\dot{A}_{\vk} +\delta \omega_k^2 A_{\vk}+\int^t_{t_0}  \Sigma(k,t-t')\,e^{i\omega_k(t-t')}\,A_{\vk}(t')\,dt' =0\,, \label{alfaeqn}  \\
&    &  \ddot{A}^*_{\vk}+2i\omega_k\,\dot{A}^*_{\vk} + \delta \omega_k^2 \, A^*_{\vk}+\int^t_{t_0}  \Sigma(k,t-t')\,e^{-i\omega_k(t-t')}\,A^*_{\vk}(t')\,dt'   = 0 \,. \label{alfastareqn}\eea Let us focus on eqn. (\ref{alfaeqn}) since (\ref{alfastareqn}) is obtained from it by $\omega_k \rightarrow -\omega_k$.

 The derivatives $\dot{A}_{\vk};\ddot{A}_{\vk}$ are written as an expansion in $g^2$ since $A_{\vk}$ is a slowly varying amplitude. In order to generate an expansion in derivatives proportional to powers of $g^2$, we write
\be \Sigma(k,t-t')\,e^{i\omega_k(t-t')} = \frac{d}{dt'}W[t;t']~~;~~W[t;t']= \int^{t'}_{t_0}  \Sigma(k,t-t'')\,e^{i\omega_k(t-t'')} \,dt'' ~~;~~ W[t;t_0]=0\,,\label{Wfun} \ee  with
\be W[t,t] = -ig^2   \int \frac{dk_0}{2\pi} \rho(k_0,k)\,\int^{t-t_0}_0 e^{-i(k_0-\omega_k)\tau}\, d\tau \,.\label{iWtt}\ee

Integrating by parts the last term in (\ref{alfaeqn}) yields
\be \ddot{A}_{\vk}-2i\omega_k\,\dot{A}_{\vk} +\delta \omega_k^2 \,A_{\vk}+ W[t,t]\,A_{\vk}(t)-  \,\int^t_{t_0}  W[t;t']\,\frac{d}{dt'}A_{\vk}(t')\,dt' =0  \,,\label{alfaw} \ee since $W[t;t'] \propto g^2$ and $dA_{\vk}(t)/dt \propto g^2$ the last term in (\ref{alfaw}) is of $\mathcal{O}(g^4)$ and will be neglected since we only keep terms up to and including $\mathcal{O}(g^2)$. Hence to $\mathcal{O}(g^2)$ the eqn. (\ref{alfaw}) simplifies to
 \be \ddot{A}_{\vk}-2i\omega_k\,\dot{A}_{\vk} +\Big[\delta \omega_k^2 + W[t,t]\Big]\,A_{\vk}(t)  =0  \,.\label{alfawg2} \ee Writing
\be A_{\vk}(t) = e^{I_{k}(t)}~~;~~ I_k(t)= g^2 \,I^{(1)}_k(t)+ g^4 \, I^{(2)}_k (t)+ \cdots \label{alfasoli}\ee
 and keeping only the leading $\mathcal{O}(g^2)$ terms, $I^{(1)}_k$ obeys the simple eqn.
 \be \ddot{I}^{(1)}_k(t) -2i\omega_k\,\dot{I}^{(1)}_k(t) = -\frac{1}{g^2}\,\Big[\delta \omega_k^2 + W[t,t]\Big]\,.\label{Idiot} \ee

 At early times $t-t_0 \simeq 1/m_a,1/T$ transient effects are expected, but we are interested in the intermediate and long time asymptotics $t-t_0 \gg 1/m_a, 1/T$, in this limit we replace
\be \int^{t-t_0}_0 e^{-i(k_0-\omega_k)\tau}\, d\tau  \rightarrow  -i \mathcal{P}\Big(\frac{1}{k_0-\omega_k} \Big) + \pi \delta(k_0-\omega_k)\,,\ee for which $W[t,t]$ becomes a constant, and choosing the counterterm
\be \delta \omega_k^2 = - g^2\,\int \mathcal{P}\Big[\frac{\rho(k_0,k)}{\omega_k-k_0}\Big]
\,\frac{dk_0}{2\pi} \,,\label{counter} \ee the solution of eqn. (\ref{alfawg2}) is
\be A_k(t) = A_k(0)\,e^{-\frac{\Gamma_k}{2}\, t}\,, \label{ampsol} \ee where
\be \Gamma_k = g^2 \, \frac{\rho(\omega_k,k)}{2\omega_k}\,. \label{gami}\ee

The finite temperature renormalized mass   is defined as the long wavelength limit of the dispersion relation, namely $m_a  = \omega_{k=0}$, therefore if follows that $\delta \omega^2_{k=0} = m^2_{0a}-m^2_a$, yielding
\be  m^2_a -m^2_{0a}  =    g^2\,\int \mathcal{P}\Big[\frac{\rho(k_0,0)}{m_a-k_0}\Big]
\,\frac{dk_0}{2\pi}\,,\label{delma}\ee and the damping rate of the misaligned axion condensate is the long wavelength limit of (\ref{gami}), i.e.
\be  \Gamma  = g^2 \,\frac{\rho(m_a,0)}{2m_a}\,, \label{quans} \ee  which are the results (\ref{misosci}-\ref{gamma}).


\begin{thebibliography}{99}

\bibitem{PQ} R. D. Peccei and H. R. Quinn,  Phys. Rev. Lett. 38, 1440
(1977), Phys. Rev. D 16, 1791 (1977).

\bibitem{weinaxion} S. Weinberg,  Phys. Rev. Lett. 40,
223 (1978).

\bibitem{wil} F. Wilczek,  Phys. Rev. Lett. 40, 279 (1978).

\bibitem{pres}  J. Preskill, M. B. Wise, and F. Wilczek,   Phys. Lett. B 120, 127 (1983).

\bibitem{abbott}  L. F. Abbott and P. Sikivie,  Phys. Lett. B 120, 133 (1983).


\bibitem{dine}  M. Dine and W. Fischler,
Phys. Lett. B 120, 137 (1983).

\bibitem{banks} T. Banks and M. Dine,   Nuclear Physics B 479,
173 (1996).

\bibitem{ringwald} A. Ringwald, Physics of the Dark Universe, 1, 116 (2012).



 \bibitem{marsh} D.J.E. Marsh,   Phys. Rept., 643, 1  (2016);  F. Chadha-Day, J. Ellis, D. J. E. Marsh, arXiv:2105.01406, D. J. E. Marsh, arXiv:1712.03018; A. Diez-Tjedor, D. J. E. Marsh, arXiv:1702.02116; J. E. Kim, D. J. E. Marsh, Phys. Rev. D93, 025027  (2016).
\bibitem{sikivie1} P. Sikivie, Rev. Mod. Phys. 93, 015004 (2021).

\bibitem{sikivie2} P. Sikivie, Lect. Notes in Physics 741, 19 (2008).






\bibitem{fuzzy} W. Hu, R. Barkana, and A. Gruzinov,   Phys. Rev. Lett., 85, 1158  (2000).

\bibitem{uldm} L. Hui, J. P. Ostriker, S. Tremaine, E. Witten, Phys. Rev. D 95, 043541 (2017).

\bibitem{quint} S. M. Carroll, Phys. Rev. Lett.81, 3067 (1998); S. Panda, Y. Sumitomo, S. P. Trivedi, Phys. Rev. D83, 083506 (2011); G. Choi, M. Suzuki, T. T. Yanagida, Phys. Lett. B 805, 135408 (2020); G. Choi, W. Lin, L. Visinelli, T. T. Yanagida, Phys. Rev. D104, L101302 (2021).

\bibitem{banik} N. Banik, A. J. Christopherson, P. Sikivie, E. M. Todarello, Phys. Rev.D95, 043542 (2017). 

 \bibitem{cast} CAST collaboration, Nature Physics, 13, 584 (2017).

\bibitem{admx} ADMX Collaboration, Phys. Rev. Lett.127, 261803 (2021). 

\bibitem{graham} P. W. Graham, I. G. Irastorza, S. K. Lamoreaux, A. Lindner, K. A. van Bibber, Ann. Rev. Nucl. Part. Sci. 65, 485 (2015).

 \bibitem{turner}  M. S. Turner,   Phys. Rev. D 28, 1243 (1983);  Phys. Rev. D 33, 889 (1986).


\bibitem{carroll} S.M. Carroll and G.B. Field, Phys. Rev. D, 43, 3789 (1991); S.M. Carroll, G.B. Field and R. Jackiw, Phys. Rev. D, 41, 1231 (1990); S. M. Carroll, G. B. Field, Phys. Rev. Lett. 79, 2394  (1997).

  \bibitem{harari}  D. Harari and P. Sikivie, Phys. Lett. 289B, 67 (1992).

  \bibitem{finelli} F. Finelli, M. Galaverni, Phys. Rev. D79, 063002 (2009).

  \bibitem{mcdonald} J. I. McDonald, L. B. Ventura, Phys. Rev. D101, 123503 (2020).

  \bibitem{beutter} M. Beutter, A. Pargner, T. Schwetz, E. Todarello,      JCAP02, 026  (2019).

  \bibitem{arza}  A. Arza, T. Schwetz, E. Todarello   JCAP 10,   013 (2020);   A. Arza, R. G. Elias,  Phys. Rev. D 97, 096005 (2018);  A. Arza,  Eur.Phys.J.C 79, 3 (2019).

 \bibitem{dashin} D. S. Lee, K. W. Ng,      	Phys.Rev. D61,    085003 (2000).

\bibitem{minami} Y. Minami, E. Komatsu, Phys. Rev. Lett.125, 221301 (2020).

 \bibitem{murai} K. Murai, F. Naokawa,T. Namikawa, E. Komatsu, arXiv:2209.07804.

 \bibitem{komatsu} E. Komatsu, Nat. Rev. Phys.4, 452 (2022).



 \bibitem{pala} P. Diego-Palazuelos \emph{et.al.} Phys. Rev. Lett. 128, 091302 (2022).

 \bibitem{eskilt} J. R. Eskilt,     Astronomy and Astrophysics 662, A10 (2022);  J. R. Eskilt and E. Komatsu,   arXiv:2205.13962.

 \bibitem{wilczekaxion} F. Wilczek, Phys. Rev. Lett. 58, 1799 (1987).

 \bibitem{wang} R. Li, J. Wang, X-L. Qi, S-C Zhang, Nature Physics 6, 284 (2010); X.-L. Qi, T. L. Hughes, S.-C. Zhang, Phys. Rev. B78, 195424 (2008). 


 \bibitem{narang} D. M. Nenno, C. A. C. Garcia, J. Gooth, C. Felser, P. Narang, Nature Reviews Physics 2, 682 (2020).

 \bibitem{gooth} J. Gooth, \emph{et.al.} Nature 575, 315 (2019).

 \bibitem{yu} J. Yu,B. J. Wieder, C.-X. Liu, Phys. Rev. B104, 174406 (2021).

 \bibitem{wilczek} L. Shaposhnikov, M. Mazanov, D. A. Bobylev, F. Wilczek, M. A. Gorlach, arXiv:2302.05111.

 \bibitem{bala}  H. S. Røising, B. Fraser, S. M. Griffin, S. Bandyopadhyay, A. Mahabir, S.-W. Cheong, A. V. Balatsky,  Phys. Rev. Research 3, 033236 (2021); A. V. Balatsky, B. Fraser, arXiv:2302.02174.

 \bibitem{binek} J. Wang, C. Lei, A. H. MacDonald, C. Binek, arXiv:1901.08536.






\bibitem{shuyang} S. Cao, D. Boyanovsky,  Phys. Rev. D 106, 123503 (2022); arXiv:2212.05161.

\bibitem{fetter} A. Fetter, J. D. Walecka, \emph{Quantum Theory of Many-Particle Systems} (Dover Publications, Inc. Mineola, New York (2003)).



  \bibitem{kms} R. Kubo, J. Phys. Soc. Jpn, \textbf{12}, 570 (1957);
 P. C. Martin and J. Schwinger, Phys. Rev.\textbf{ 115}, 1342
(1959).



 \bibitem{htl}
E. Braaten and R.D. Pisarski, Nucl. Phys. {\bf B337}, 569 (1990);
{\bf B339}, 310 (1990); R.D. Pisarski, Physica A {\bf 158}, 146
(1989); Phys. Rev. Lett. {\bf 63}, 1129 (1989); Nucl. Phys. {\bf
A525}, 175 (1991), E. Braaten, R. Pisarski, Phys. Rev. \textbf{D45}, R1827 (1992).






\bibitem{lebellac} M. Lebellac, \emph{Thermal Field Theory}, (Cambridge Monographs on Mathematical Physics), (Cambridge University Press, United Kingdom, 1996).


\end{thebibliography}
\end{document}